\documentclass[12pt, onecolumn, superscriptaddress,notitlepage]{revtex4-1}
\usepackage{color}
\usepackage{amsmath,amssymb,graphicx,bm,float}

\begin{document}
\title{High-efficiency single-photon generation via large-scale active time multiplexing}
\author{Fumihiro Kaneda} 
\affiliation{Department of Physics, University of Illinois at Urbana-Champaign, Urbana, IL 61801, USA}
\author{Paul G. Kwiat} 
\affiliation{Department of Physics, University of Illinois at Urbana-Champaign, Urbana, IL 61801, USA}

\maketitle

\textbf{
On-demand generation of indistinguishable single- and multi-photon states is a key technology for scaling up optical quantum information and communication applications. 
Nonlinear parametric photon-pair sources and heralded single-photon sources (HSPSs) had been the most standard resource of quantum information applications for decades \cite{Pan:2012kv}. 
However, the intrinsic uncertainty of the produced number of photon pairs in such sources is a critical drawback that prevents on-demand photon-pair and heralded single-photon generation. 
Here we demonstrate large-scale time multiplexing of indistinguishable heralded single photons, employing a low-loss HSPS and adjustable delay line. 
We observed 66.7$\pm 2.4$\% presence probability of single-photon states collected into a single-mode optical fiber by multiplexing 40 periodic time bins of heralded single photons. 
To our knowledge, this is the highest fiber-coupled single-photon probability achieved to date. 
A high indistinguishability ($\sim$90\%) of our time-multiplexed photons has also been confirmed. 
We also experimentally investigate trade-off relations of single-photon probability and unwanted multi-photon contribution by using different pump powers for a HSPS. 
Our results demonstrate that low-loss, large-scale multiplexing can realize highly efficient single-photon generation as well as highly scalable multi-photon generation from inefficient HSPSs. 
We predict that our large-scale time multiplexing will pave the way toward generation of $> 30$ coincident photons with unprecedented efficiencies, enabling a new frontier in optical quantum information processing. 
}

In the past two decades, photon-pair sources based on spontaneous parametric downconversion (SPDC) and spontaneous four-wave mixing (SFWM) have been used for many groundbreaking quantum information experiments. 
However, it is very difficult to further scale up quantum information and communication applications by simply using multiple photon-pair sources, since photon pairs cannot be generated deterministically; 
for a mean number of photon pairs per pump pulse $\mu$, the generation probability of $k$ photon pairs is given by $\mu^k / (\mu +1 )^{k+1}$. 
Therefore, the single-pair generation probability peaks at only 25\% due to the non-negligible likelihood ($\sim \mu^k$) of unwanted zero- and multiple-pair generations. 
For example, a recent ten-photon experiment \cite{Wang:2016dk} using 5 SPDC sources needed to keep $\mu < 0.05$ to suppress the multi-pair emissions, resulting in a ten-photon coincidence generation rate of only $\sim 0.1$ per second and a final detection rate of only several events per hour.

\begin{tiny}
\begin{figure}[t!]
 \includegraphics[width=0.98\columnwidth,clip]{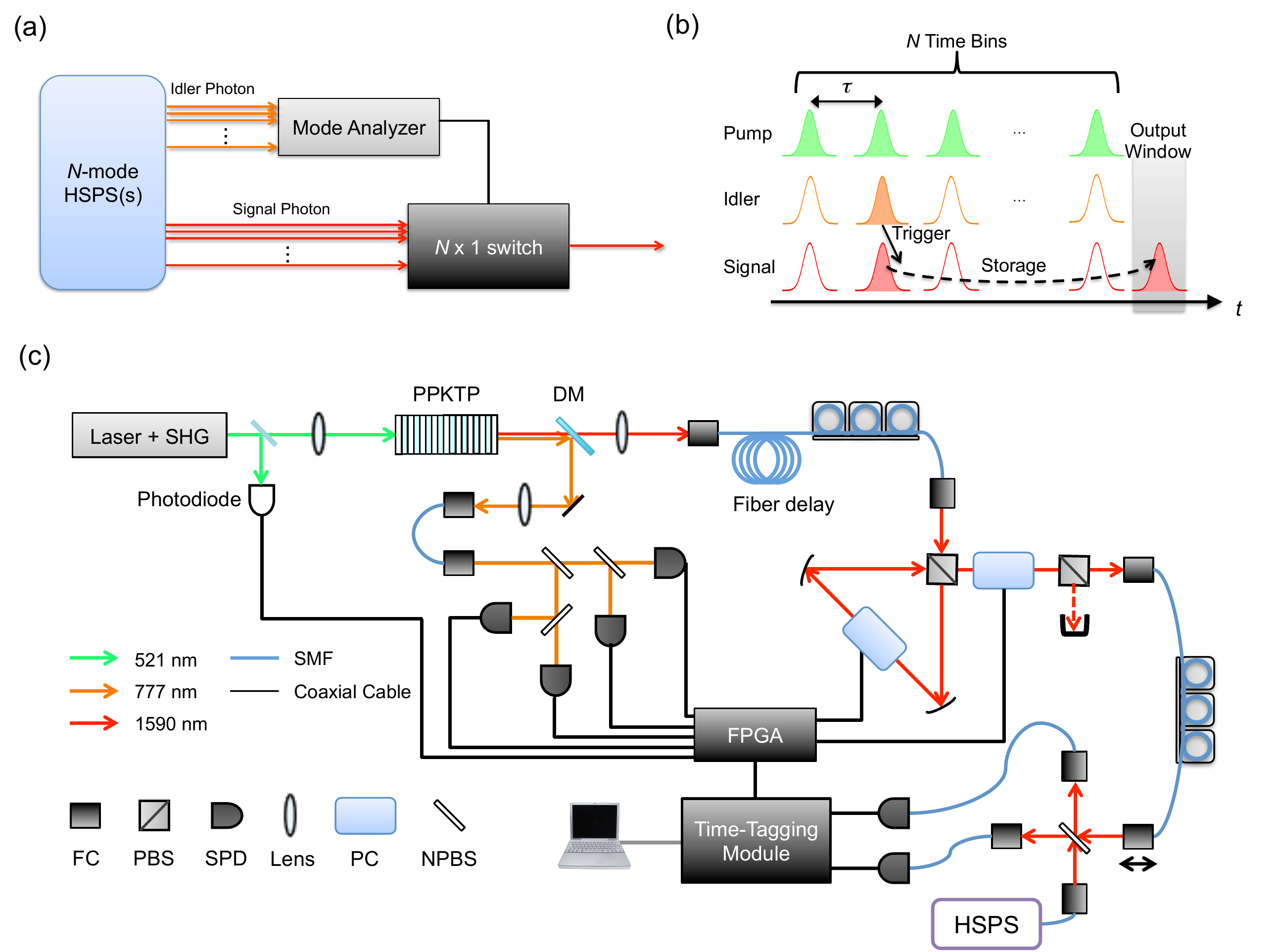}
 \caption{Multiplexed heralded single-photon sources (HSPSs) and our implementation. 
 (a) Simplified diagram of a general multiplexed HSPS. Multi-mode SPDC source(s) probabilistically generate photon pairs in which a signal-photon state is correlated to its twin idler-photon state. 
 According to a mode analysis of an idler photon, an adaptive $N \times 1$ optical switch converts a signal photon state to a predetermined output mode, e.g., time bin. (b) Timing diagram of our time-multiplexed scheme. 
 Our HSPS pumped with a period $\tau$ (probabilistically) generates photons in $N$ different time bins. 
 An adjustable delay line can delay signal photons for an arbitrary integer multiples of $\tau$ so that any initial time bin state of a heralded photon is converted to a fixed output time bin. 
 (c) Schematic diagram of our experimental setup. 
 SHG, second harmonic generation; PPKTP, periodically-poled potassium titanyl phosphate crystal; DM, dichroic mirror; FC, fiber coupler; PBS, polarizing beam splitter; SPD, single-photon detector; PC, Pockels cell; NPBS, nonpolarizing beam splitter; SMF, single-mode fiber; and FPGA, field-programmable gate array. See Methods for experimental details. }
\label{setup}
\end{figure}
\end{tiny}

In 2002, Pitmann and co-workers \cite{Pittman:2002dx} and Migdall and co-workers \cite{migdall:2002hk} independently proposed ``multiplexing'' as a technique to overcome the probabilistic nature of SPDC sources. 
In general multiplexing methods (see Fig. \ref{setup}a), a twin photon (signal photon) generated by multi-mode, probabilistic SPDC processes is rerouted to a single mode by adaptive optical switches controlled in accordance with a mode analysis of the other twin photon (idler photon), whose mode is correlated (or entangled) to that of the ``heralded'' signal photon. 
In this case, the single-photon generation probability is no longer constrained by the 25\% limit, and can reach as high as a multiplexed heralding probability 
\begin{align}
P_{H} = 1-(1 - p)^N,
\end{align} 
where $N$ is the number of multiplexed modes and $p$ is the probability of a trigger detector signal per mode, i.e., approximately the product of $\mu$ and the trigger system detection efficiency $\eta_T$ for one SPDC mode. 
Thus, for a sufficiently large number of multiplexed modes, one can achieve pseudo-deterministic generation of heralding signals and thereby heralded single photons. 
Since the proposals in 2002 this promising method has been theoretically analyzed and extended \cite{Jeffrey:2004ky, McCusker:2009ci, Glebov:2013hr, Mower:2011kp, Adam:2014ix, Mazzarella:2013cj, Bonneau:2015ho, GimenoSegovia:2017de}.

For practical implementations of multiplexed HSPSs, there are two challenges in addition to large-scale multiplexing: first, losses in both HSPS and optical switches should be very low, because lost single-photon states, unlike classical states of light, cannot be restored. 
The other challenge is to generate \textit{indistinguishable} photons, necessary to achieve high-visibility multi-photon interference, central to most photonic quantum-gate operations \cite{Knill:2001is,Kok:2007ep}. 
Therefore, HSPSs and optical switches also need to generate and maintain pure single-photon states in each multiplexed mode and then convert those to an identical pure state. 
Some experimental attempts have demonstrated either large-scale multiplexing \cite{Kaneda:2015dn,Mendoza:2016gr}, low-loss photon generation/rerouting \cite{Kaneda:2015dn,Ma:2011in}, or indistinguishable single-photon generation \cite{Ma:2011in, Xiong:2016bv, FrancisJones:2016je}, but none of previous experiments simultaneously achieves these three key requirements.

Here we demonstrate the first large-scale and low-loss multiplexing to produce heralded single photons with very high probability and indistinguishability. 
In order to achieve such a high-quality single-photon source, we implemented a time-multiplexing system (extending the method proposed by Pittman and co-workers \cite{Pittman:2002dx}) as shown in Fig. \ref{setup}b,c. 
With our method one can implement a large-scale multiplexing with only one HSPS and adjustable delay line (which in general works as a quantum memory \cite{Kaneda:2017gp}), while other demonstrated methods such as spatial-multiplexing \cite{migdall:2002hk} and frequency-multiplexing \cite{Puigibert:2017fh, Joshi:2017vj} methods require a number of HSPSs and/or optical switches. 
In our time-multiplexing scheme, a HSPS pumped by $N$ sequential pump pulses with a period $\tau$ (= 10 ns) probabilistically generates photon pairs. 
An adjustable delay line triggered by a heralding signal stores heralded photons for an arbitrary integer multiple of $\tau$. 
A stored photon is released at a certain predetermined output time bin regardless its birth time bin, and thereby $N$ time-bin modes of heralded single photons are multiplexed into the single output time bin. 
We implemented this scheme with high-quality components: our heralded photons at 1590 nm have 88\% coupling efficiency into a first collection single-mode fiber and 91\% spectral indistinguishability \cite{Kaneda:2016fh}. 
An adjustable delay line consisting of a very low loss optical switch (a Pockels cell, PC) and polarizing beam splitter (PBS) and built with careful spatial-mode management, has only 1.2\% loss per cycle. 
Moreover, to maintain the temporal indistinguishability of multiplexed single photons, the cycle length is matched to $\tau$ within 0.01 ps, much less than the $6.1$-ps pulse duration of heralded single photons.  
The delay line's group-velocity dispersion, which can also make single photons distinguishable in terms of storage cycles, is negligible for the 0.8-nm bandwidth of our single photons.

\begin{tiny}
\begin{figure}[t!]
\begin{center}
 \includegraphics[width=1\columnwidth,clip]{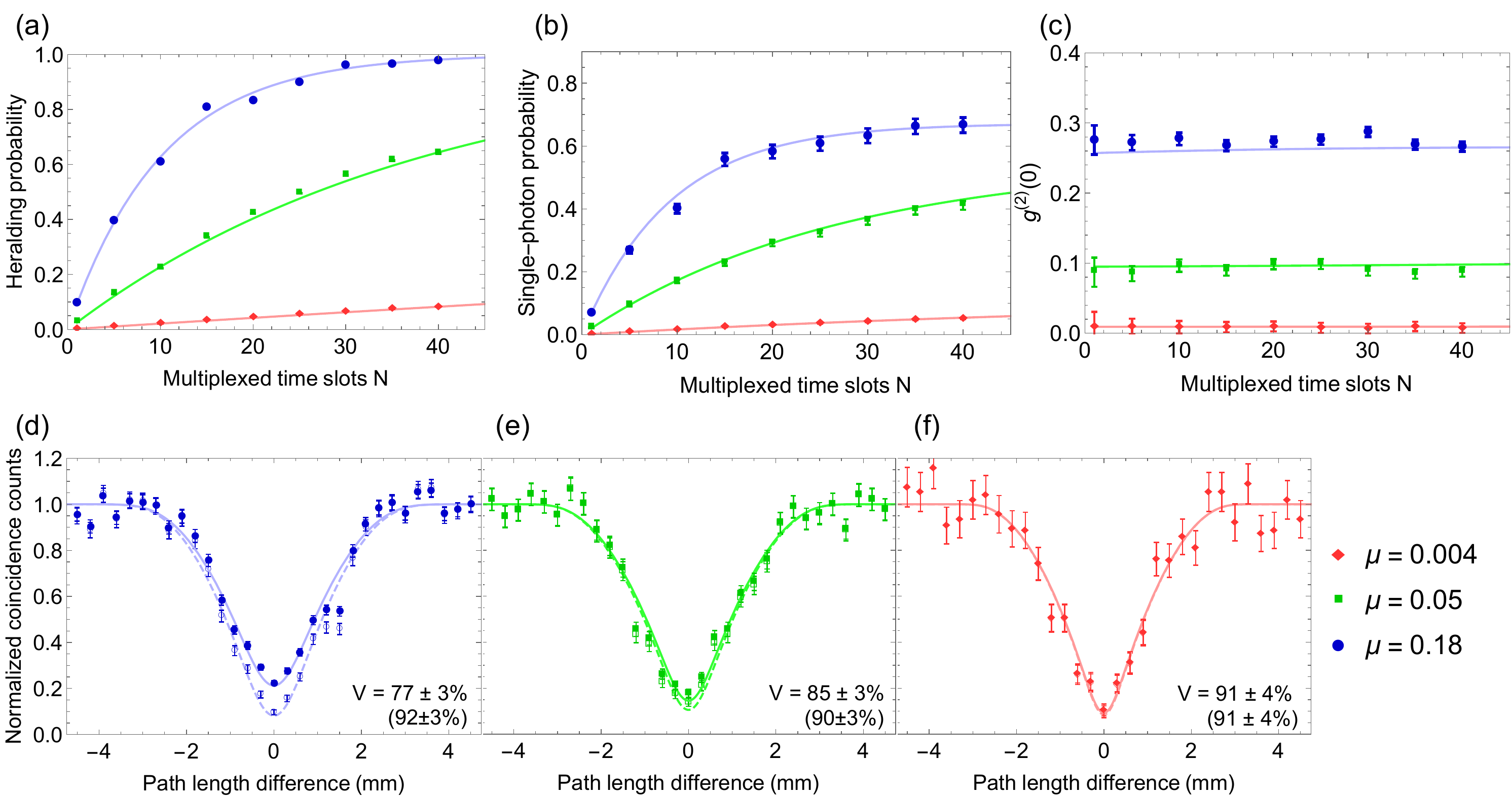}
   \caption{Experimental results. (a) heralding signal probability $P_H$, (b) single-photon probability $P_1$, and (c) second-order auto-correlation function $g^{(2)} (t =0)$ versus number of multiplexed time bins $N$. $P_H$ and $P_1$ are significantly enhanced as $N$, while $g^{(2)} (0)$ is approximately unchanged for all different mean photon numbers $\mu$. 
(d,e,f) Observed HOMI for synchronized photons with $N = 40$. 
Empty circles, squares, and diamonds in (d,e,f) show data points after subtracting accidental coincidences. 
Solid and dashed lines are the best fit theoretical curves \cite{Kaneda:2016fh} for raw coincidences and those after subtracting accidental coincidences, respectively. 
$V$ denotes the interference visibility without (with) subtracting accidental coincidences. 
Error bars are estimated by Poissonian photon counting statistics. 
}
\label{result1}
\end{center}
\end{figure}
\end{tiny}

We characterized our time-multiplexed photons using a setup with a nonpolarizing beam splitter (NPBS) and two single-photon detectors (SPDs) to measure single-photon counts and unwanted multi-photon contributions simultaneously. 
The multiplexed heralding probability shown in Fig. \ref{result1}a is estimated as $P_H =  H/R$, where $H$ is the heralding signal rate and $R = 500$ kHz is the repetition rate of the multiplexing process. 
Figure \ref {result1}b shows the multiplexed single-photon probability estimated by 
\begin{align}
P_1 = \frac{1}{R \eta} \left\{S_1+S_2 - C\left(\frac{4}{\eta} -1\right)\right\}, 
\label{P1}
\end{align} 
where $S_i$ is a single count rate of Detector $i$, $C$ is a coincidence count rate between Detector 1 and 2, and $\eta = 0.426$ is the net transmission of the optics from the second collection fiber (after the delay line) to SPDs. 
In Eq. (\ref{P1}) $C$ is included to correct for multi-photon contributions in $S_1$ and $S_2$ (see Methods for details). 
Our source was tested for 3 different mean photon numbers per pulse: 
For $\mu = 0.18$, we observed nearly saturated $P_H$ with $N = 40$ time bins. 
With this condition, we observed $P_1 = 0.667 \pm 0.024$, corresponding to $10\times$ enhancement over the non-multiplexed case (for $N = 1$). 
To our knowledge, this is the highest single-photon probability after single-mode-fiber coupling (some semiconductor single-photon sources \cite{Somaschi:2016ej, Loredo:2016kl, Wang:2017jw} have demonstrated higher single-photon extraction probabilities after a first collection lens, but poor coupling into a single-mode fiber, and therefore substantially poorer potential performance in any applications that requires photons in a single spatial mode).  
With $\mu = 0.05$ and $0.004$ and $N = 40$, we observed $(P_H,P_1) = (0.639,   0.412 \pm 0.013), (0.082,   0.051 \pm 0.002)$, respectively. 
Although the heralding probabilities did not reach saturation for these lower mean photon numbers, the enhancement factors are respectively $19 \times$ and $28 \times$ for $\mu = 0.05$ and $0.004$, much higher than the one for $\mu = 0.20$ in our source. 

Unwanted multi-photon contributions can be quantified by the second-order auto-correlation function $g^{(2)} (t =0)$, which can be estimated by
\begin{align}
g^{(2)} (t =0)= \frac{CR}{S_1 S_2}. 
\label{g2}
\end{align}
For $\mu = 0.18$, 0.05, and 0.004, we observed respectively $g^{(2)} (0)$ $\sim 0.27, \sim 0.09$, and $\sim 0.009$, all of which were approximately constant versus $N$ (see Fig. \ref{result1}c).  
This indicates that the time-multiplexing technique successfully enhances the single-photon generation probability without increasing the multi-photon noise relative to the single-photon fraction. 
The $g^{(2)} (0)$ increases with $\mu$ due to higher multi-photon noise, and therefore $g^{(2)} (0)$ and $P_1$ are in trade-off relations. 
However, note that $g^{(2)} (0)$ can be further suppressed without dropping $P_1$, by introducing high-efficiency photon-number-resolving detection (e.g., with superconducting detectors \cite{Marsili:2013fs}) for the idler mode to herald one and only one signal-photon emission event.

We directly measured the indistinguishability of the time-multiplexed photons by Hong-Ou-Mandel interference (HOMI) \cite{Hong:1987vi}. 
For this measurement we prepared an additional non-multiplexed HSPS whose heralded single photons were coupled to the second input port of the NPBS so they could be interfered with the time-multiplexed photons (see Fig. \ref{setup}c). 
Our observed HOMI dips with $N = 40$ are shown in Fig. \ref{result1}d,e,f.  
The respective estimated visibilities with the best fit theoretical curves \cite{Kaneda:2016fh} for $\mu = 0.18, 0.05$, and 0.004 were $V = 77\pm3\%, 85\pm3\%$, and $91\pm4\%$ with raw coincidence count rates, i.e., without accidental subtraction. 
Our observed visibility is lower for higher $\mu$ due to higher multi-photon noise as shown in Fig. \ref{result1}b. 
However, since those visibilities after subtracting background counts are $\sim90\%$ for all mean photon numbers,  
we conclude that the spectral and temporal indistinguishability of our heralded photons is well maintained after the multiplexing operation via the adjustable delay line. 

\begin{table}[t!]
\caption{Comparison of performances of single-photon sources. MUX-HSPS, multiplexed-HSPS; QD quantum dot; $I$, indistinguishability; $C_{M} = P_1^M R$, predicted $M$-fold coincidence generation rate assuming that $M$ independent sources can be prepared and synchronously operated. Note that $P_1$ is the probability of preparing a single photon that is coupled into a single-mode fiber. For sources reported for different experimental parameters, results with conditions demonstrating the highest $P_1$ are shown.}
\begin{tabular}{   cccccccc }      \hline
  Ref.                                        &   Method                               &   $R$    &   $P_1$                    &     $g^{(2)} (t =0)$   &  $I$                         &   $C_{10}$ (/s)           &  $C_{30}$ (/s)         \\  \hline 
 \cite{Wang:2016dk}    & SPDC       & $\sim170$ MHz &  $\sim0.04 $  & $\sim0.1$  & $0.91$& $\sim10^{-1}$& $\sim10^{-19}$  \\   
\cite{Ma:2011in}    & MUX-HSPS                             & 80 MHz &  $\sim0.001$  & $\sim0.5$  & $0.89$ & $\sim10^{-22} $& $\sim10^{-82}$  \\ 
\cite{Kaneda:2015dn}   & MUX-HSPS                           & 50 kHz &  $0.386$  & $0.48$  & $\sim 0.05$& $\sim4$& $\sim10^{-8}$  \\  
  \cite{Xiong:2016bv}   & MUX-HSPS                            & 10 MHz &  $\sim0.0024$  & $\sim0.2$               & 0.91   & $\sim10^{-19}$& $\sim10^{-72}$  \\
 \cite{Somaschi:2016ej} &        QD      & 82 MHz &  $\sim 0.001$  & $ 0.0028$  & $ 0.996$& $\sim10^{-22} $& $\sim10^{-82}$  \\
 \cite{Loredo:2016kl} &        QD      & 80 MHz &  $0.14$  & $ 0.013$  & $0.7$& $\sim0.1 $& $\sim10^{-18}$  \\
 \cite{Wang:2017jw}    & QD                & 76 MHz &  $0.337$  & $0.027$  & $0.93$& $\sim10^3$& $\sim10^{-7}$  \\   \hline
  This work          & MUX-HSPS                            & 500 kHz &  $0.667$  & $0.269$  & $0.91$& $\sim10^4$& $\sim 1$  \\ 
  Possible     improvement                     & MUX-HSPS                            & 5 MHz &  $0.75$  & $0.05$  & $0.98$& $\sim10^5$& $\sim 10^3$   \\ \hline
\label{comparison}
\end{tabular}
\end{table}

A comparison of our source with other representative single-photon sources is shown in Table \ref{comparison}. 
Our source significantly outperforms all other state-of-the-art sources in $P_1$, and has a spectral indistinguishability comparable to the best indistinguishable photon sources. 
Although the $g^{(2)} (0)$ for our source is worse than semiconductor sources, using high-efficiency trigger SPDs would make it comparable, as mentioned above. 
We can also consider how well our method would perform at producing multiple individual photons. 
$C_{M} =  P_1^M R$ denotes an $M$-photon coincidence generation rate with the assumption that $M$ identical sources are prepared and operated synchronously. 
Despite its relatively low repetition rate of 500 kHz, our time-multiplexed source can generate ten-photon coincidences with $\sim 10^4/$s success rates, 5 orders of magnitude better than a recent ten-photon experiment with non-multiplexed SPDC sources \cite{Wang:2016dk}. 
Note that further improvement of our time-multiplexed HSPS is possible with the-state-of-the-art technologies (see Supplementary Information); with feasible upgrades we expect 30-photon coincidence rates $\sim 10^4$ /s, and even 50-photon coincidence rate can be $> 1$ /s.

In conclusion, we have demonstrated large-scale time multiplexing for efficient and indistinguishable single-photon generation.  
Our developed HSPS and adjustable delay line make it possible to multiplex up to 40 time-bin modes of heralded single photons with very low loss and high indistinguishability. 
Consequently, our time-multiplexed HSPS has significantly better single-photon emission probability than those in previous demonstrations, and expected orders-of-magnitude better multi-photon coincidence rates. 
We anticipate that this time-multiplexed source with such unprecedented efficiencies will be an optimal resource for large-scale photonic quantum computation systems, particularly for quantum walk \cite{Childs:2013hh} and boson sampling \cite{Aaronson:2011ja} systems, to demonstrate ``quantum supremacy'' over classical computation systems.

 \section*{Acknowledgements}
Funding for this work has been provided by NSF Grant No. PHY 12-12439 and PHY 15-20991, US Army ARO DURIP Grant No. W911NF-12-1-0562, ARO Grant No. W911NF-13-1-0402, and US Navy ONR MURI Grant No. N00014-17-1-2286.

\section*{Methods}
\subsection*{Heralded single-photon source}
A 20-mm-long periodically-poled potassium titanyl phosphate (PPKTP) crystal pumped by a frequency-doubled Yb laser ($\lambda = 521$ nm, $\tau = 10.0 $ ns) generates collinear photon pairs via SPDC process. 
Idler photons at 777 nm are detected by a cascade of four Si-avalanche-photodiode detectors, each of which has a $\sim$62\% detection efficiency. 
The detector cascade allows us to reduce the effect of each SPD's saturation and to resolve the approximate photon number in a time bin. 
Signal photons at 1590 nm are sent to an adjustable delay line via a (fixed) fiber delay line.  
With a group-velocity-matching condition in the SPDC process, 91\% spectral indistinguishability of heralded single photons is achieved without spectral filtering \cite{Kaneda:2016fh}. 
Transmission probabilities for idler and signal photons after a first collection fiber are 84\% and 88\%, respectively. 
Similar performance is observed in another ``non-multiplexed'' HSPS used for HOMI measurements. 
For this latter non-multiplexed HSPS we used $\mu = 0.008$, so that unwanted multi-photon states degrade the HOMI visibilities only $\sim1$\%. 

\subsection*{Fixed pre-delay line}
The heralded photons are first directed into a 100-m fiber delay line, which holds the photons for $\sim$500 ns. 
This compensates for the electronic latencies ($\sim 100$ ns from a trigger photon to firing the PC); the rest of the delay ($\sim 400$ ns) allows us to select the latest heralded time bin for up to $N =40$.
Storing only the ``latest-born'' heralded photon minimizes the number of storage cycles and the associated loss \cite{Kaneda:2015dn}. 
After the fixed fiber delay the photons are directed into the adjustable delay.

\subsection*{Adjustable delay line}
Our adjustable delay line consists of a 10-ns delay loop, custom Brewster-angled PBS, and PC comprising a pair of rubidium titanyl phosphate (RTP) crystals. 
A field-programmable gate array (FPGA) module processes input signals from four trigger SPDs, triggering the PC. 
When a heralded photon enters into the delay line through the PBS, the PC is activated, rotating that photon's polarization by 90$^\circ$ to store and delay it in the loop. 
After delaying for the necessary integer multiples of $\tau = 10$ ns, the stored photon is released by a second switching of the PC. 
The single-pass cavity loss is 1.2\%, corresponding to a photon $1/e$ lifetime of $\sim$ 830 ns (= 83 cycles).  
The slight loss is due to the loss in the PC ($0.8$\%) and two concave mirrors (0.2\% for each). 
The group-velocity dispersion of the adjustable delay line is $\sim$$1.2\times 10^{-3}$ ps$^2$ per cycle, negligibly small compared to the photon coherence time (6.1 ps). 
Thus, the cycle-dependent chromatic dispersion, which could degrade spectral indistinguishability of the synchronized photons, is negligible for up to $N = 40$. 
Our delay line is built with high-mechanical-stability optics mounts in a temperature-stabilized laboratory, and has a small long-term cycle length drift ($\sim$0.01 ps per hour), much less than the 6.1-ps single-photon pulse duration. 
A PC and PBS after the delay line comprise an optical shutter that only transmits photons heralded in an allowed output time bin. 

\subsection*{Estimation of $P_1$}
We estimated a single-photon probability $P_1$ assuming that the probability of more than two photons is negligibly small; a multiplexed $k$-photon probability $P_{k}$ for $k \geq3$ is expected $< 1.5$\% of $P_1$ for $\mu = 0.18$ in our experiment. 
We modeled our SPDs as bucket-detectors that discriminate between zero or one-or-more photons. 
The NPBS's transmission/reflection ratio is also assumed to be 1. 
With those assumptions, single and coincidence count rates are given by 
\begin{align}
S_i &= \frac{P_1 R  \eta}{2 } +P_2 R\left\{ \frac{\eta}{2}+\frac{\eta(2- \eta)}{4}\right\}, 
\label{S}
\end{align} 
\begin{align}
C &=\frac{P_2R  \eta^2}{2 }, 
\label{C}
\end{align} 
where $P_2$ is the probability of a two-photon state after multiplexing. 
The second and third terms in Eq. (\ref{S}) are the respective detector click probabilities that one and two of two photons arrive at Detector $i$. 
Equation (\ref{P1}) is obtained by substituting Eq. (\ref{C}) into Eq. (\ref{S}).

\end{document}